\documentclass[aps,prb]{revtex4}
\usepackage{amsmath,amssymb}
\usepackage{graphicx}
\usepackage{dcolumn}
\usepackage{bm}
\usepackage{color}
\usepackage{relsize}
\usepackage{mathtools}
\newcommand{\mbb}{\mathbb}
\newcommand{\mc}{\mathcal}

\newcommand{\tet}{\texttt}
\newcommand{\pr}{\partial}

\begin{document}
\title{Developing a semiclassical Wentzel–Kramers–Brillouin theory for $\alpha-\mc{T}_3$ model}
\author{Kathy Blaise$^{1}$, Chinedu Ejiogu$^{1}$, 
Andrii Iurov$^{1}$\footnote{E-mail contact: aiurov@mec.cuny.edu, theorist.physics@gmail.com
},
Liubov Zhemchuzhna$^{1,2}$,
Godfrey Gumbs$^{2,3}$, 
and
Danhong Huang$^{4,5}$
}

\affiliation{
$^{1}$Department of Physics and Computer Science, Medgar Evers College of City University of New York, Brooklyn, NY 11225, USA\\ 
$^{2}$Department of Physics and Astronomy, Hunter College of the City University of New York, 695 Park Avenue, New York, New York 10065, USA\\ 
$^{3}$Donostia International Physics Center (DIPC), P de Manuel Lardizabal, 4, 20018 San Sebastian, Basque Country, Spain\\ 
$^{4}$Space Vehicles Directorate, US Air Force Research Laboratory, Kirtland Air Force Base, New Mexico 87117, USA\\ 
$^{5}$Center for High Technology Materials, University of New Mexico, 1313 Goddard SE, Albuquerque, New Mexico, 87106, USA\\
}

\date{\today}

\begin{abstract}
We have developed a complete semiclassical Wentzel-Kramers-Brillouin (WKB) theory for $\alpha-\mc{T}_3$ model which describes a wide 
class of existing pseudospin-1 Dirac cone materials. By expanding the sought wave functions in a series over the powers of Planck constant 
$\hbar$, we have obtained the leading order expansion term which is the key quantity required for calculating the electronic and transport properties of a semiclassical electron in $\alpha-\mc{T}_3$. We have derived the transport equations connecting each two consecutive orders of the wave function expansion and solved them to obtained the first order WKB wavefunction. We have also discussed the applicability of the obtained approximation and how these results could be used to investigate various tunneling and transport properties of $\alpha-\mc{T}_3$ materials with non-trivial potential profiles. Our results could be also helpful for constructing electronics devices and transistors based on innovative flat-band Dirac materials.  
\end{abstract}
\maketitle

\section{Introduction} 
\label{sec1}

The quantum-mechanical description of realistic electronic behavior in various existing materials is predictably a lot more complicated than any classical analysis of such phenomena. Therefore, it is important to build a semiclassical approximation for high-energy and fast-moving electrons with a simplified description which would still lead to a physically correct picture. Therefore, creating a semiclassical WKB theory is one of the most important ingredients for any new type of Hamiltonian and we believe that doing so for a newly discovered $\alpha-\mc{T}_3$ and a dice lattice would be deemed as a crucial advancement in addition to the current research being performed in the field low-dimensional condensed matter physics.\,\,\cite{vogl, wkbg1, wkbg2, weekes2021generalized, zalip, zalip2, wkb1} 

\par 

Generally, the idea of a WKB approximation is based on finding an approximated solution of a differential equation with spatially dependent coefficients in the case when one of those them (which physically corresponds to the external potential $V(x)$) is varying considerable slower than the others which related to a the rate of change of the wavelength for our electron. Therefore, the obtained solutions is interpreted as a rapidly oscillating a quantum state modulated by a non-essential change of the external potential. The WKB solution is obtained by expanding the wavefunction over the powers of Planck constant $\hbar$ and in this respect is viewed as semiclassical. It was previously obtained for all Dirac materials, specifically graphene and gapped graphene. WKB theory has numerous applications, specifically, in calculating the electron tunneling for nontrivial potential barriers,\,\cite{sonin, fanwar} investigating into the resonant tunneling and resonant scattering, trapped and localized electronic states.\,\cite{yeke, xu1, gangaraj2020topological}  

\par 

Specifically, the present work is intended to develop a complete semiclassical WKB theory for the so-called $\alpha-\mc{T}_3$ model.\,\cite{berc, dora1, vidal1, vidal2001disorder,thesis} This is a general model for a large group of recently discovered and innovative low-dimensional materials known by the presence of a flat band in their energy dispersions in addition to a regular Dirac cone.  This flat band is stable and remains in the presence of a number of external perturbations, including magnetic and electric fields, point defects and off-resonance dressing fields.\,\cite{ourpeculiar, dey1, dey2, prx1, ourPh, kisrep, kiMain, ourJAP2017ph, fnam, kibis2017all,kibis2022floquet}

\par 

The atomic structure of $\alpha-\mc{T}_3$ materials is represented by a regular hexagon lattice similar to graphene with an additional atom in the center of each hexagon. This buildup leads to the existence of extra electron hopping coefficient associated with the hub atom which results in a dispersionless band in the low-energy bandstructure of these materials. The effect and the "weight" of the electron transition from and into this flat band is defined by a quantum phase $\phi$ and ranges from 0 (graphene with a completely uncoupled flat band) and a dice lattice with $\phi = \pi/4$ and the strongest possible effect of the flat band.  These highly unusual energy dispersions result in unique and truly fascinating electronic, \,\cite{cunha2022tunneling, Malc01, ourml, fuf1, Dutta1, ourplay, cunha2021band, iurov2021tailoring} transport,\,\cite{alphaDice, alphaKlein, wa20, c0} optical, \,\cite{han2022optical,car1, our20, opt1} and magnetic, \,\cite{nic2,  nic1, piech1, piech2, tutul2, tutul1,balassis2020magnetoplasmons} properties of these innovative materials, which are very different from those in graphene.\,\cite{neto} Together with tilted Dirac cone materials\,\cite{tilt1, tilt2, tilt3} and topological Dirac semimetals\,\cite{isl1, isl2}, $\alpha-\mc{T}_3$ materials are considered the most promising and innovating low-dimensional lattices at the preset time.

\par  
At this moment, a large number of $\alpha-\mc{T}_3$ materials and dice lattices have been fabricated. These include a trilayer
SrTiO$_3$/SrIrO$_3$/SrTiO$_3$, \,\cite{R110, dice1} Hg$_{1-x}$Cd$_x$ quantum well,\,\cite{orlita2014observation, Malc01} Josephson arrays,\,\cite{han2022optical, R200} Leib and Kagome optical lattices and optical waveguides.\,\cite{R117, R116, R115, Dan6, Dan2, Dan7, Dan1, R133, Dan5}, In$_{0.53}$Ga$_{0.47}$As/InP semiconducting layers\,\cite{nnfl} and many-many other existing  materials. A comprehensive review of all Dirac materials with a flat band can be found in Ref.\,[\onlinecite{Add1}].

\par  

To sum up, we would like to refer a reader to a complete and timely review of all existing flat-band structures published in 2019.\,\cite{Add1} Pseudospin-1 Dirac-Weyl Hamiltonian used to model the electronic behavior for $\alpha-\mc{T}_3$ model describes the properties of all these materials with a certain degree of similarity.\,\cite{ournew, LiuM}

Both theoretical and experimental investigation into electronic properties has become an important direction of nowadays condensed matter physics, nearly all possible issues related to the electronic phenomena in $\alpha-\mc{T}_3$ have been now addressed. Thus, we feel that building a semiclassical WKB theory for such new type of materials is one of the crucial problems which still needs to be solved.

\medskip
\par
The rest of the present paper is organized in the following way. First, in Sec.\,\ref{sec2} we review the low-energy Hamiltonian, bandstructure and the electronic states for various types of $\alpha-\mc{T}_3$ lattices, including graphene and a dice lattice. Section \,\ref{sec3}, which is the key section of this paper, deals with calculating the semiclassical action and the longitudinal electron momentum, finding 
the leading order wavefunctions, deriving the most general system of transport equations which connect each to consequent orders of the $\hbar$-expansion of the wave function, calculating the first order wave function and verifying the applicability of the WKB approximation for our model Hamiltonian. In Sec. \,\ref{sec4}, we briefly discuss the applications of the obtained semiclassical wave function to studying the 
electron tunneling and estimating the transmission coefficient for various non-trivial potential potential profiles. Finally, we provide some 
concluding remarks and outlook in Sec. \,\ref{sec5}.

\section{Electronic states in pseudospin-1 Dirac cone materials}
\label{sec2}

\medskip
\par
We are going to build and use a semiclassical WKB approximation for $\alpha-\mc{T}_3$ model with the low-energy electronic states 
described by a pseudospin-1 low-energy Dirac-Weyl Hamiltonian

\begin{equation}
\label{H0}
\hat{\mc{H}}_{\alpha} (\mbox{\boldmath$k$} \, \vert \, \tau, \phi) = \hbar v_F 
\left[
\begin{array}{ccc}
0 & k^\tau_- \, \cos \phi &  0 \\
k^\tau_+ \, \cos \phi & 0 & k^\tau_- \, \sin \phi \\
0 & k^\tau_+ \, \sin \phi & 0
\end{array}
\right] \, ,  
\end{equation}

where phase $\phi$ is obtained from the relative hopping parameter $\alpha$ as $\alpha = \tan \phi$ and $k^\tau_\pm = \tau k_x \pm i k_y$ depend on the valley index $\tau = \pm 1$ corresponding to $K$ and $K'$ valleys.  

\par 
For $\phi=0$, Hamiltonian \eqref{H0} is obviously reduced to that of graphene. The opposite limiting case $\phi=\pi/4$ defined as a dice lattice, corresponds to the strongest effect of the presence of the additional hub atom and is always selected as a separate case if of the highest importance. We will explicitly provide all our results for the dice lattice.

\medskip
\par
If the electron is also subjected to a finite position-dependent  potential $V(x)$, such as a square barrier $V(x)=V_0 \Theta(x)
\Theta(W_B - x)$ or a barrier with non-piecewise-uniform potential profile, the longitudinal momentum $p_x(x)$ also depends on $x$ and Hamiltonian \eqref{H0} could be rewritten as 

\begin{equation}
\label{HamG}
\hat{\mc{H}}_{\alpha} (\mbox{\boldmath$k$} \, \vert \, \tau, \phi)  = v_F \, \hat{\mbox{\boldmath$\Sigma$}}(\phi) \cdot \left\{ -i \hbar\mbox{\boldmath$\nabla$}_\tau \right\} + V(x) \, \hat{\Sigma}^{(3)}_{0} \ ,
\end{equation}
in terms of $\phi$-dependent $3 \times 3$ Pauli matrices $\hat{\Sigma}^{(3)}(\phi)= \left\{\hat{\Sigma}^{(3)}_x(\phi),\,\hat{\Sigma}^{(3)}_y(\phi) \right\}$

\begin{equation}
\label{Sxp}
\hat{\Sigma}_x(\phi) = \left[
\begin{array}{ccc}
 0 & \cos \phi & 0 \\
 \cos \phi & 0 & \sin \phi \\
 0 & \sin \phi & 0
\end{array}
\right] \ ,
\end{equation}

and

\begin{equation}
\label{Syp}
\hat{\Sigma}_y(\phi) = i \,\left[
\begin{array}{ccc}
 0 & -\cos \phi & 0 \\
 \cos \phi & 0 & -\sin \phi \\
 0 & \sin \phi & 0
\end{array}
\right] \ .
\end{equation}

A $3\times 3$ unit matrix $\hat{\Sigma}^{(3)}_{0}$ 

\begin{equation}
\label{S13}
\hat{\Sigma}^{(3)}_{0} = \left[
\begin{array}{ccc}
 1 & 0 & 0 \\
 0 & 1 & 0 \\
 0 & 0 & 1
\end{array}
\right] \ ,
\end{equation}
which also enters Eq.~\eqref{HamG}, does not depend on phase $\phi$, and $\mbox{\boldmath$\nabla$}_\tau=\{ \tau \pr/\pr x,\,\pr/\pr y \}$. 
For a dice lattice with $\phi=\pi/4$ matrices Eqs.\,\eqref{Sxp} and  \eqref{Syp} are simplified to well-known $3 \times 3$ Pauli matrices 

\begin{equation}
\hat{\Sigma}^{(3)}_{x} = \frac{1}{\sqrt{2}} \, \left[
\begin{array}{ccc}
 0 & 1 & 0 \\
 1 & 0 & 1 \\
 0 & 1 & 0
\end{array}
\right] \ .
\label{sig1}
\end{equation}

and

\begin{equation}
\hat{\Sigma}^{(3)}_{y} = \frac{i}{\sqrt{2}} \, \left[
\begin{array}{ccc}
 0 & -1 & 0 \\
 1 & 0 & -1 \\
 0 & 1 & 0
\end{array}
\right] \ .
\label{sig2}
\end{equation}

For a finite bandgap the third Pauli matrix  

\begin{equation}
\label{szdice}
\hat{\Sigma}^{(3)}_{z} = \left[
\begin{array}{ccc}
 1 & 0 & 0 \\
 0 & 0 & 0 \\
 0 & 0 & -1
\end{array}
\right] \ , 
\end{equation}
could be utilized.\,\cite{Gusgap, gorb}

\medskip
\par
The energy eigenvalues of Hamiltonian \,\eqref{H0} are

\begin{equation}
\label{vc}
\varepsilon^{\gamma=\pm 1}_{\tau, \, \phi}(\mbox{\boldmath$k$}) =  \gamma \hbar v_F k
\end{equation}
where $\gamma = - 1$ ($\gamma = + 1$) corresponds to the valance (conduction) band, and 
$\gamma = 0$ leads to a dispersionless solution 

\begin{equation}
\label{fb}
\varepsilon^{\gamma=0}_{\tau, \, \phi}(\mbox{\boldmath$k$}) = 0\ .
\end{equation}
known as a "flat band". All three eigenenergies of \eqref{H0} do not depend on phase $\phi$.

for the remaining flat (or dispersionless) band. 
Here, all three bands in Eqs.\,\eqref{vc} and \eqref{fb} do not show any dependence on phase $\phi$ (or parameter $\alpha$).

The wave functions corresponding to the valence and conduction bands in Eq.\,\eqref{vc} are given as

\begin{equation}
\label{Eig1}
\Psi^{\gamma=\pm 1}_{\tau, \, \phi}(\mbox{\boldmath$k$})  = \frac{1}{\sqrt{2}} \left[
\begin{array}{c}
\tau \cos \phi \,\, \tet{e}^{- i \tau \theta_{ \bf k}}  \\
\gamma \\
\tau \sin \phi \,\, \tet{e}^{+ i \tau \theta_{ \bf k}} 
\end{array}
\right]\ ,
\end{equation}
where $\theta_{\bf k} = \arctan (k_y/k_x)$ is the angle of wave vector $\mbox{\boldmath$k$} = \{k_x, k_y\}$ made with the $x$-axis. The remaining wave function attributed to the flat band is 

\begin{equation}
\label{Eig2}
\Psi^{\gamma=0}_{\tau, \, \phi}(\mbox{\boldmath$k$}) = \left[
\begin{array}{c}
\sin \phi \,\, \tet{e}^{- i \tau \theta_{\bf k}}  \\
0 \\
- \cos \phi \,\, \tet{e}^{+ i \tau \theta_{\bf k}} 
\end{array}
\right]\ . 
\end{equation}

Here, we would like to indicate that the energy bands in Eqs.\,\eqref{vc} and \eqref{fb}, as well as the wave functions in Eqs.\,\eqref{Eig1} and \eqref{Eig2}, are obtained for a spatially-uniform potential independent of position coordinates $x$ and $y$.
\medskip

\begin{figure} 
\centering
\includegraphics[width=0.6\textwidth]{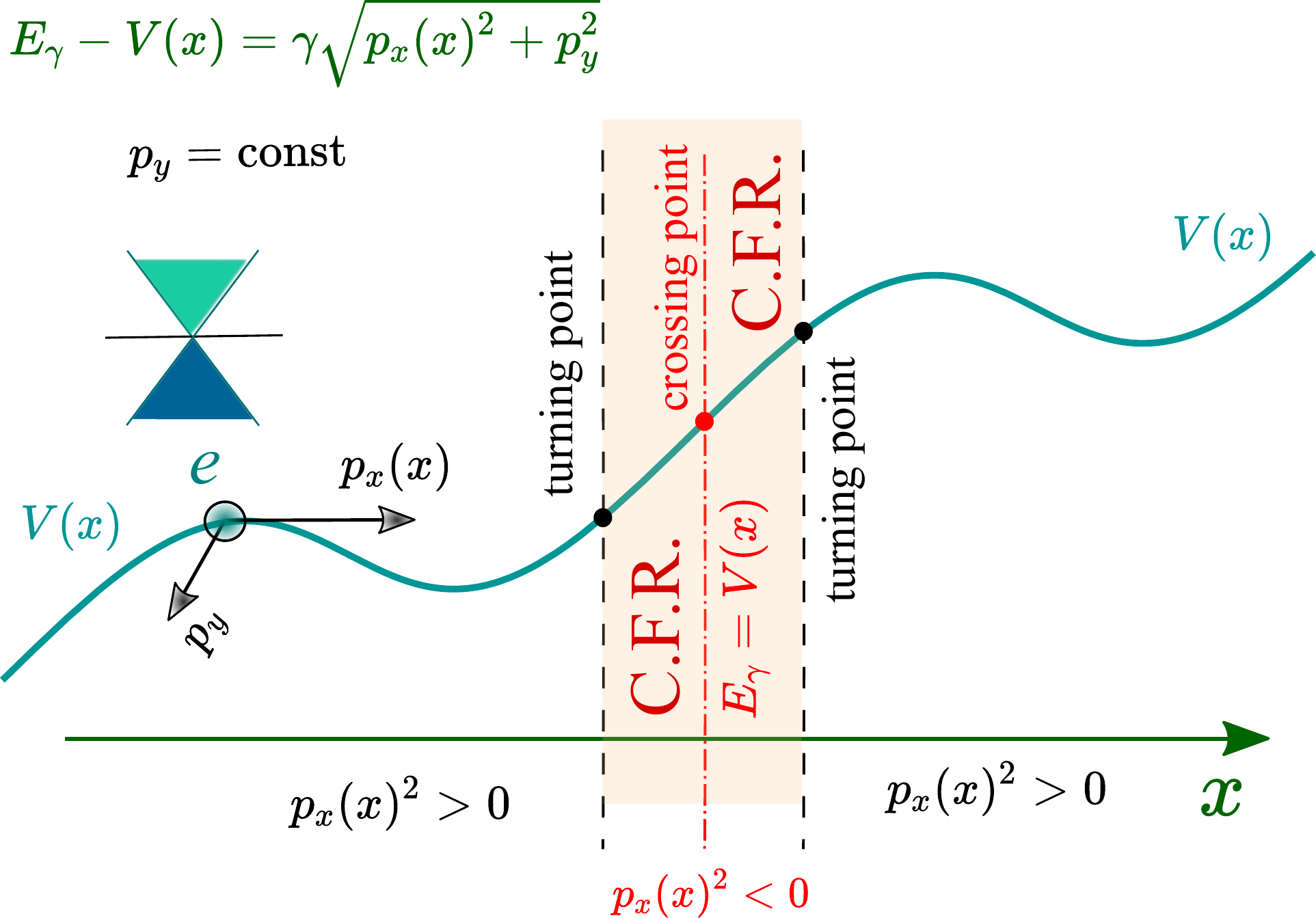}
\caption{(Color online) Schematics for a Dirac electron moving in a varying potential $V(x)$. The longitudinal momentum for such electron also 
depends on its position $x$ while the transverse momentum $k_y$ is conserved. We also demonstrate classically forbidden or classically inaccessible regions with $p_x(x)^2 < 0$ and its boundaries known as the turning points. The crossing point is a location of the particle for which $V(x) = E_\gamma$ so that the transition between the electron and hole states is observed.}
\label{FIG:1}
\end{figure}

For a dice lattice, wave functions \eqref{Eig1} and \eqref{Eig2} are as follows:

\begin{equation}
\label{Eig1d}
\Psi^{\gamma=\pm 1}_{\tau, \, \phi=\pi/4}(\mbox{\boldmath$k$})  = \frac{1}{2} \, \left[
\begin{array}{c}
\tet{e}^{- i \tau \theta_{ \bf k}}  \\
\sqrt{2} \, \tau \gamma \\
\tet{e}^{+ i \tau \theta_{ \bf k}} 
\end{array}
\right]\ ,
\end{equation}

and 

\begin{equation}
\label{Eig2d}
\Psi^{\gamma=0}_{\tau, \, \phi=\pi/4}(\mbox{\boldmath$k$}) = \frac{1}{\sqrt{2}} \, \left[
\begin{array}{c}
 \tet{e}^{- i \tau \theta_{\bf k}}  \\
0 \\
- \tet{e}^{+ i \tau \theta_{\bf k}} 
\end{array}
\right]\ , 
\end{equation}
in which two components have equal magnitudes.

\section{Semi-classical WKB model and solution} 
\label{sec3}

Here, we intent to derive the so-called transport equations connecting different orders of the series expansion of the WKB wave function in powers of $\hbar$, This is a set of the most important differential equation which allow for calculation the wave function with any required 
precision.  The most important is the principal zero-order wave function which requires the knowledge of the semi-classical action being the 
spatial derivative of the position-dependent longitudinal electron momentum of our electron. 

\par 
We also provide the general solutions for our transport equations, calculate the next (first-order) expansion term of the electron wave functions and demonstrate the applicability conditions for WKB approximation $\psi_1 \ll \psi_0$. This is the central section of our paper containing its most crucial findings.

\subsection{General solution and action}

Our main consideration is an electron moving in $x$-dependent potential $V(x)$ so that the longitudinal momentum $p_x(x)$ is
also non-uniform along the $x$-direction. The wave function is presented as $\Psi(x,y) \backsim \psi(x)\,\tet{e}^{i k_y \, y}$
due to the fact that the translational symmetry in the $y$-direction is still preserved. 

\medskip  

Hamiltonian in Eq.\,\eqref{HamG} is now given by 

\begin{eqnarray}
\label{H02}
&&\hat{\mc{H}}_{\alpha} (x, k_y \, \vert \, \tau) = 
\left\{
\begin{array}{ccc}
V(x) &  v_F \, \cos \phi\,\Xi(x)_\tau^{-} &  0 \\
 v_F \, \cos \phi\,\Xi(x)_\tau^{+} & V(x) &  v_F \, \sin \phi\,\Xi(x)_\tau^{-}\\
0 & v_F \, \sin \phi\,\Xi(x)_\tau^{+} & V(x)
\end{array}
\right\} \, , \\
\nonumber
&& \Xi(x)_\tau^{\pm} = - i \hbar \, \tau \, \frac{\pr}{\pr x} \pm i p_y \, ,
\end{eqnarray}
which reflect the remaining translational symmetry in the $y-$direction. Also, we used a $p_y = \hbar k_y$ notation. We describe 
this situation in Fig.~\ref{FIG:1}. If the potential energy of the particle is increased, its longitudinal momentum will become 
zero at some points which are defined as "turning points". These points specify the classically forbidding regions - the areas
with imaginary longitudinal momentum $p_x(x)^2 < 0$. The point where $E_\gamma = V(x)$ and the particle has zero energy is the 
crossing point between the electron and hole states. In contrast to graphene, the electron momentum in $\alpha-\mc{T}_3$ at such 
points cannot be determined due to its infinite degeneracy for $E_\gamma - V(x) = 0$.

\medskip 
Since we are planning to build up an approximation and estimate each of its terms, it seems reasonable to rewrite our initial 
eigenvalue equation in a dimensionless form. A similar approach was adopted in Refs.\,[\onlinecite{zalip}] and [\onlinecite{weekes2021generalized}]. First, we 
divide each of equations \eqref{H02} by the unit of energy $V_0$ as $E \longrightarrow E/V_0$ and $V(x) \longrightarrow V(x)/V_0$. $V_0$ could be a maximum height of the potential barrier or the potential energy of our particle at an any fixed point. The length is renormalized over the barrier width $W_B$ as $x \longrightarrow x / W_B$. Thus, the new dimensionless momentum is obtained as $p_{x,y} \longrightarrow  v_y p_{x,y}/V_0$ and, most importantly, our expansion parameter the Planck constant is modified as  

\begin{equation}
\hbar \longrightarrow \left(\frac{v_F}{W_B \, V_0}\right)\ \hbar \ .
\end{equation}

The spatial derivative would change as $\pr/\pr x \rightarrow \pr / (W_B \, \pr x)$. We feel reluctant to connect the energy unit through 
the inverse length (a unit of momentum) is it would bring another term dependent on a small parameter $\hbar$ into our calculation. Thus, 
the kinetic energies of the incoming particles are assumed to be quasi-classical, e.i., having much larger values than the Fermi energy of 
our system.

\medskip 
\par
Our main eigenvalue equation is now written as 

\begin{eqnarray}
\label{new1}
&&\hat{\mc{H}}_{\alpha} (x, p_y \, \vert \, \tau) \, \Psi_\alpha^{\,\gamma}(x, p_y \, \vert \, \tau) = E_\gamma \, \Psi^{\,\gamma}
(x, p_y \, \vert \,  \phi, \tau)
\end{eqnarray}

where the wavefunction $\Psi_\alpha^{\,\gamma}(x, p_y \, \vert \, \tau)$ is expressed as 

\begin{eqnarray}
&&\Psi_\alpha^{\,\gamma}(x, p_y \, \vert \, \tau) = \psi_\alpha ^{\,\gamma}(x \, \vert \, \phi, \tau) \, \tet{exp}\left(\frac{i}{\hbar} \,
p_y y \right) =  \left[
\begin{array}{c}
\phi_A (x \, \vert \, \phi, \tau) \\
\phi_H (x) \\
\phi_B (x \, \vert \, \phi, \tau)
\end{array}
\right]
\tet{exp}\left(\frac{i}{\hbar} \, p_y y \right)\ .
\label{new2}
\end{eqnarray}

Hamiltonian in Eq.\,\eqref{new1} now becomes 

\begin{eqnarray}
\nonumber
\hat{\mc{H}}_{\alpha} (x, p_y \, \vert \, \tau) &=& 
\hat{\Sigma}^{(3)}_{0} \, V(x) + \hat{\Sigma}^{(3)}_{x}(\phi) \, \left[
- i \hbar \, \tau \, \frac{\pr}{\pr x}  \right] + \hat{\Sigma}^{(3)}_{y}(\phi)  p_y =  \\
&=& \left\{
\begin{array}{ccc}
V(x) &  v_F \, \Xi(x)_\tau^{-} \, \cos\phi  & 0 \\
v_F \, \Xi(x)_\tau^{+} \, \, \cos\phi &  V(x) &  v_F \, \Xi(x)_\tau^{-} \, \sin\phi \\
0 & v_F \, \Xi(x)_\tau^{+} \, \sin\phi  &  V(x)
\end{array}
\right\} \, .
\label{h_new} 
\end{eqnarray} 

\medskip 
\par 

A standard Wentzel–Kramers–Brillouin semiclassical approach is developed by expanding the unknown wave function in a series over the powers of 
$\hbar$

\begin{eqnarray}
\label{expand}
&& \Psi(\xi, p_y \, \vert \, \phi, \tau)  = \tet{exp} \left\{\frac{i}{\hbar} \, \mbb{S}(x) \right\}\, \sum \limits_{\lambda = 0}^{\infty} (- i \hbar)^\lambda \, \Psi_\lambda (x) = \\
\nonumber
&& = \tet{exp} \left\{ \frac{i}{\hbar} \, \mbb{S}(x) \right\}\, \left[\Psi_0 (x) - i \hbar \, \Psi_1 (x) - \hbar^2 \, \Psi_2 (x)  + \cdots \, \right]\ ,
\end{eqnarray}
where $\mbb{S}(x)$ is the semi-classical action, which we still need to calculate. 

\par 
The problem of building a semiclassical approximation for an $x-$dependent potential will be solve once we obtain the WKB wavefunctions or 
all the terms of expansion \eqref{expand}. Therefore, we need to derive a set of differential equations which connect all consecutive terms of expansion \eqref{expand}. These equations, referred to as the {\it transport equations}, are the key relations in any semiclassical theory. 

\par 

Explicitly, the transport equation for $\alpha-\mc{T}_3$ is derived as 

\begin{equation}
\label{transeq}
\hat{\Sigma}^{(3)}_{x} (\phi) \, \left\{\, \frac{\pr}{\pr x} \,  \Psi_{\lambda}(x, p_y \, \vert \, \phi, \tau)  \, \right\}
- \frac{1}{\sqrt{2}} \, \hat{\mbb{O}}_{\,T} (x, p_y \, \vert \, \phi \, \tau) \, \Psi_{\lambda+1} (x, p_y \, \vert \, \phi, \tau) 
= 0 \ ,
\end{equation}
where $\lambda = 0, \,1, \,2, \, 3, \,\cdots$ and

\begin{equation}
\Psi_{\lambda <  0}(\xi, p_y \, \vert \, \phi, \tau) \equiv 0 \, . 
\end{equation}
Here, the transport operator $\hat{\mbb{O}}_{\,T} (\xi, p_y \, \vert \, \phi, \, \tau)$, connecting consequent terms of expansion in Eq.\,\eqref{expand}, is easily found to be

\begin{equation}
\label{to1}
\hat{\mbb{O}}_{\,T} (x, p_y \, \vert \, \phi, \tau) = 
\left\{
\begin{array}{ccc}
\nu(x) & \Xi_{\,\mbb{S},\tau}^{(-)} \, \cos \phi  & 0 \\
 \Xi_{\,\mbb{S},\tau}^{(+)} \, \cos \phi & \nu(x)
&  \Xi_{\,\mbb{S},\tau}^{(-)} \, \sin \phi \\
0 & \Xi_{\,\mbb{S},\tau}^{(+)} \, \sin \phi & \nu(x)
\end{array}
\right\} \, , 
\end{equation}
where 

\begin{eqnarray}
&& \Xi_{\,\mbb{S},\tau}^{(\pm)} = \tau \, \frac{\pr\mbb{S}(x)}{\partial x} \pm i p_y \, , \\
\nonumber 
&& \nu(x) = V(x) - E_\gamma \, . 
\end{eqnarray}
Specifically, by setting $\lambda = - 1$ in Eq.\,\eqref{transeq}, we obtain a simplified equation for the zeroth order wavefunction  

\begin{equation}
\label{zeroord}
\hat{\mbb{O}}_{\,T} (x, p_y \, \vert \, \phi, \tau) \, \Psi_0 (x, p_y \, \vert \, \phi, \, \tau) = 0 \ ,
\end{equation}

which is a linear and homogeneous algebraic system; a non-trivial solution for such such system could be found only 
if its determinant equals to zero resulting in the following equation

\begin{equation}
- \left[E_\gamma - V(x) \right]\, \left\{
 \, \left(\frac{\pr \, \mbb{S}(x)}{\pr x} \right)^2 + p_y^2 - \left[E_\gamma - V(x) \right]^2 \,
\right\} = 0 \ ,
\label{new-4}
\end{equation}
which is independent of $\phi$. For a non-uniform potential profile $V(x)$, condition $E_\gamma = V(x)$ could not 
be always kept, apart from a finite number of isolated points. Therefore, the classical action $\mbb{S}(x)$

\begin{eqnarray}
\label{sp}
&& \mbb{S}(x) = \mbb{S}(x_0) + \int\limits_{x_0}^{x} p_x(\eta) \, d \eta \ , \\
\label{sp2} 
&& p_x(x) = \pm \sqrt{ \left[ E_\gamma - V(x) \right]^2 - p_y^2} 
\end{eqnarray}
is expressed through the position-dependent longitudinal momentum $p_x(x)$. Transverse momentum $p_y$ always remains the same 
due to the translational symmetry.

\subsection{Leading order wave function}

The principal step in building a semiclassical approximation is calculating the leading order (zero-order) electron wavefunction  
$\Psi_0 (x,p_y \, \vert \, \phi, \tau)$ which has the following form

\begin{eqnarray}
&& \Psi_0 (x, p_y \, \vert \, \phi, \tau) = \Phi_0 (x, p_y \, \vert \, \phi, \tau) \, ,  \\
\label{zo}
&& \Phi_0 (x, p_y \, \vert \, \phi, \tau) = \left[
\begin{array}{c}
\varphi_A^{(0)} (x \, \vert \, \phi, \tau) \\
\varphi_H^{(0)} (x) \\
\varphi_B^{(0)} (x \, \vert \, \phi, \tau)
\end{array}
\right]\ .
\end{eqnarray}
Using Eqs.~\eqref{to1} and \eqref{sp}, we can connect the components of wave function \eqref{zo}

\begin{eqnarray}
\nu(x)\,\varphi_A^{(0)}(x) + \cos\phi \, \left[\tau p_x(x)-i p_y\right]\,\varphi_H^{(0)}(x)&=&0\ ,\\
\sin \phi\,\left[\tau p_x(x)+i p_y \right]\,\varphi_H^{(0)}(x) + \nu(x)\,\varphi_B^{(0)}(x)&=&0\ ,
\end{eqnarray}
which now accepts the following form

\begin{eqnarray}
\label{wf0theta}
&& \Psi_0 (x \, \vert \, \phi, \, \tau) = \left[
\begin{array}{c}
\cos \phi \,\, \Theta(x \, \vert \, \tau) \\
-1 \\
\sin \phi \,\, \Theta^{\star}(x \, \vert \, \tau)
\end{array}
\right]\,\varphi_H^{(0)} (x) \ , \\
\label{theta}
&& \Theta(x \, \vert \, \tau) = \frac{1}{\nu(x)} \, \left[ \tau p_x(x) - i p_y \right] = 
- \tau \, \tet{exp}\left[ - i \tau \, \theta_{\bf k}(x) \right]\, , \\
\nonumber 
&& \Theta^{\star}(x \, \vert \, \tau) \xRightarrow[]{i \rightarrow -i} \Theta(x \, \vert \, \tau)
\end{eqnarray}
Angle $\theta_{\bf k}(x) = \tan^{-1} [k_x(x)/k_y]$ associated with the wavevector ${\bf k} = \{k_x(x), k_y\}$ now depends on 
coordinate $x$, e.i., does not remain the same as the electron moves in potential profile $V(x)$. Another $x-$dependent scalar function 
$\varphi_H^{(0)} (x)$ from \eqref{wf0theta} remains undefined at the moment, we will need to determine it from general transport 
equation \eqref{transeq}. 

\medskip
Next, we connect the zero and the first order wavefunctions $\Psi_{1} (x, p_y \, \vert \, \phi, \tau)$ and 
$\Psi_{1} (x, p_y \, \vert \, \phi, \tau)$ by taking $\lambda = 0$ in Eq.\,\eqref{transeq}

\begin{equation}
\label{transeq1}
\hat{\mbb{O}}_{\,T} (x, p_y \, \vert \, \phi, \, \tau) \, \Psi_{1} (x, p_y \, \vert \, \phi, \tau) = 
\sqrt{2} \, \hat{\Sigma}^{(3)}_{x} (\phi)  \, \frac{\pr}{\pr x} \,  \Psi_{0}(x, p_y \, \vert \, \phi, \tau)\ ,
\end{equation} 

Wave function $\Psi_{1} (x, p_y \, \vert \, \phi, \tau)$, as any other vector in three-dimensional Hilbert space could be 
decomposed as a linear combination of three orthogonal basis vectors 
$ \big \vert \mbox{\boldmath$v$}_1 \bigr>$, $ \big \vert \mbox{\boldmath$v$}_2 \bigr>$ and $\big \vert \mbox{\boldmath$v$}_3 \bigr>$. 

The first basis vector $ \big \vert \mbox{\boldmath$v$}_1 \bigr>$ is equivalent to a part of $\Psi_0 (\xi \, \vert \, \phi,  \tau)$ in Eq.\,\eqref{wf0theta}  

\begin{equation}
\big \vert \mbox{\boldmath$v$}_1(x \, \vert \, \phi,  \tau) \bigr>  = \left[
\begin{array}{c}
\cos \phi \,\, \Theta(x \, \vert \, \tau) \\
-1 \\
\sin \phi \,\, \Theta^{\star}(x \, \vert \, \tau)
\end{array}
\right]\ .
\label{vec1}
\end{equation}
This means that $\hat{\mbb{O}}_{\,T} (x, p_y \, \vert \, \phi, \tau)  \big \vert \mbox{\boldmath$v$}_1 \bigr> = 0$. We also choose 
two remaining orthogonal vectors $\big \vert \mbox{\boldmath$v$}_2(\xi\,\vert\,\phi,\tau) \bigr>$  and $\big \vert \mbox{\boldmath$v$}_3(\xi\,\vert\,\phi,\tau) \bigr>$ in the following way

\begin{eqnarray}
\big \vert \mbox{\boldmath$v$}_2(x \, \vert \, \phi,  \tau) \bigr>  &=& \left[
\begin{array}{c}
\cos \phi \,\, \Theta(x \, \vert \, \tau) \\
+1 \\
\sin \phi \,\, \Theta^{\star}(x \, \vert \, \tau)
\end{array}
\right]\ ,\\
\big \vert \mbox{\boldmath$v$}_3(x \, \vert \, \phi,  \tau) \bigr> &=& \left[
\begin{array}{c}
\sin \phi \,\, \Theta(x \, \vert \, \tau) \\
0 \\
- \cos \phi \,\, \Theta^{\star}(x \, \vert \, \tau)
\end{array}
\right] \ . 
\end{eqnarray}
Vector $\Psi_{1} (x, p_y \, \vert \, \phi, \tau)$ is now expanded in terms of the basis $\big \vert \mbox{\boldmath$v$}_1 \bigr> $,
 $\big \vert \mbox{\boldmath$v$}_2 \bigr>$ and  $\big \vert \mbox{\boldmath$v$}_3 \bigr> $

\begin{equation}
\label{3exp}
\Psi_{1} (x, p_y \, \vert \, \phi, \tau) = \varphi_H^{(1,1)} (x)\, \big \vert\mbox{\boldmath$v$}_1(x\,\vert\,\phi,\tau)\bigr> + 
\varphi_H^{(1,2)} (x)\,\big \vert\mbox{\boldmath$v$}_2(x\,\vert\,\phi,\tau)\bigr>+\varphi_H^{(1,3)} (x)\,\big \vert\mbox{\boldmath$v$}_3(x\,\vert\,\phi,\tau)\bigr>\ . 
\end{equation}
\medskip

Now, substituting Eq.\,\eqref{3exp} into Eq.\,\eqref{transeq1}, we find

\begin{eqnarray}
\nonumber
&&\bigr< \mbox{\boldmath$v$}_1(x\,\vert\,\phi,\tau)\,\big \vert \, \hat{\mbb{O}}_{\,T} (x, p_y \, \vert \, \phi, \tau)\,\Psi_{1} (x, p_y \, \vert \, \phi, \tau)\bigr>\\
&=&\bigr< \mbox{\boldmath$v$}_1(x\,\vert\,\phi,\tau)\,\big \vert \, \hat{\mbb{O}}_{\,T} (x, p_y \, \vert \, \phi,\tau)\,\big\vert \left\{\big\vert\,
\varphi_H^{(0)} (x)\, \big \vert\mbox{\boldmath$v$}_1(x\,\vert\,\phi,\tau)\bigr> + 
\varphi_H^{(1,2)} (x)\,\big \vert\mbox{\boldmath$v$}_2(x,\vert\,\phi,\tau)\bigr>+\varphi_H^{(1,3)} (x)\,\big \vert\mbox{\boldmath$v$}_3(x\,\vert\,\phi,\tau)\bigr>\right\}\bigr> 
= 0\ \ \ \ \ \ \ \
\end{eqnarray}
since transport operator $\hat{\mbb{O}}_{\,T} (x, p_y \, \vert \, \phi, \tau)$ is Hermitian.

The first order transport equation Eq.\,\eqref{transeq12} becomes

\begin{equation}
\label{sig}
\bigr< \mbox{\boldmath$v$}_1(x\,\vert\,\phi,\tau)\, \big \vert \,  \hat{\Sigma}^{(3)}_{x} (\phi)  \, \frac{\pr}{\pr x} \,  \Psi_{0}(x, p_y \, \vert \, \phi, \tau)\,\bigr > = 
\bigr< \mbox{\boldmath$v$}_1(x\,\vert\,\phi,\tau)\, \big \vert \,  \hat{\Sigma}^{(3)}_{x} (\phi)  \, \frac{\pr}{\pr x} \, \left\{ \,
  \big \vert \,  \mbox{\boldmath$v$}_1(x \,\vert\, \phi, \tau)  \bigr> \, \varphi_H^{\,(0)} (x) \, \right\} \,
\bigr>= 0\ 
\end{equation}

could be written as 

\begin{eqnarray}
\label{mainEq}
&& \left[ 
\Theta(x \, \vert \,\tau) + \Theta^{\star}(x \, \vert \,\tau)
\right] 
\, \frac{\pr \, \varphi_H^{(0)} (x)}{\pr x} + \Gamma_{\alpha}(x \, \vert \,\tau)   \, \varphi_H^{(0)} (x) = 0 \, , 
\end{eqnarray}

where

\begin{equation}
\Gamma_{\alpha}(x \, \vert \,\tau) = \frac{\pr \, \Theta(x \, \vert \,\tau) }{\pr x} + \sin^2 \phi \, \frac{\pr}{\pr x} 
\left[ \Theta^{\star} (x \, \vert \,\tau) - \Theta(x \, \vert \,\tau)  \right] \, .
\end{equation}
The phase-dependent term could be simply rewritten through relative hopping parameter $\alpha$ as $ \sin^2 \phi = (1 + \alpha^2)^{-1}$.
Also, it worth noting that for a dice lattice with $\phi=\pi/4$ function $\Gamma_{\alpha = 1}(x \, \vert \,\tau) = \pr/ \pr x [\Theta(x \, \vert \,\tau) +\Theta^{\star}(x \, \vert \,\tau)]$ becomes exactly symmetric for $\Theta(x \, \vert \,\tau) \leftrightarrow \Theta^{\star}(x \, \vert \,\tau)$. For $\phi = 0$ we see the same equation as was derived for graphene in Ref.~[\onlinecite{zalip}].

\par
Eq.~\eqref{mainEq} has a general solution

\begin{equation}
\label{genint}
\varphi_H^{(0)} (\xi, \phi) = c_0 \, \tet{exp} \left\{
- \int \left[ \Theta(x \, \vert \,\tau) + \Theta^{\star}(x \, \vert \,\tau) \right]^{-1}
\Gamma_{\alpha}(\zeta \, \vert \,\tau)  \, d\zeta 
\right\} \ .
\end{equation}

Now we need to directly evaluate the following quantities: 

\begin{eqnarray}
\nonumber
&& \Theta(x \, \vert \,\tau) = \frac{p_{\tau,-}}{\nu(x)} = - \frac{\tau p_x - i p_y}{p} \, , \\
\nonumber
&& \Theta(x \, \vert \,\tau)^{\star} = \frac{p_{\tau,+}}{\nu(x)} = - \frac{\tau p_x + i p_y}{p} \, , \\
\nonumber
&& \left[ \Theta(x \, \vert \,\tau) + \Theta(x \, \vert \,\tau)^{\star} \right]^{-1} = - \frac{\tau \, p(x)}{2 p_x(x)} \, , \\
\nonumber 
&& \frac{\pr \, \Theta(x \, \vert \,\tau) }{\pr x} = - \frac{i \tau p_y}{p_{\tau,+}^{3/2} \, p_{\tau,-}^{1/2}} \, \frac{\pr \, p_x(x)}{\pr x} \, , \\
\label{estim}
&& \frac{\pr \, \Theta(x \, \vert \,\tau)^{\star} }{\pr x} = \frac{i \tau p_y}{p_{\tau,+}^{1/2} \, p_{\tau,-}^{3/2}} \, \frac{\pr \, p_x(x)}{\pr x} 
\end{eqnarray}

Using identities \eqref{estim}, equation \eqref{genint} is evaluated as 

\begin{equation}
\label{fr2}
\varphi_H^{(0)} (\xi, \phi) = - \sqrt{\frac{p_x(x) + i p_y}{p_x(x)}} + \frac{i}{2} \sin^2 \phi \, \tan^{-1} \left[ - i \,
\frac{p_x(x)}{p_y} \right] \, . 
\end{equation}

We note that equation \eqref{fr2} is valid up to a normalization constant and up to a phase factor. Thus, our result for 
gapless graphene with $\phi = 0$ differs by a complex phase $p_x(x) \pm i p_y$ compared to Ref.\,[\onlinecite{zalip}].  

\medskip

For a dice lattice, we obtain  

\begin{equation}
\label{vph}
\varphi_H^{(0)}(\xi) =  \left\{ \frac{p^2_x(x)+p_y^2}{p^2_x(x)}\right\}^{1/4} = \left\{ 1 +
\left[\frac{p_y}{p_x(x)} \right]^2 \, \right\}^{1/4}  \, .
\end{equation}
Now the leading order wave function $\Psi_0(\xi\,\vert\,\phi,\tau)$ in Eq.\,\eqref{wf0theta} is determined completely, including the phase differences and the spatial dependence of each its component. Apart from that, all the next orders of the wave function expansion \eqref{expand}
could be now calculated, as we will do below for the first order $\Psi_1(\xi\,\vert\,\phi,\tau)$.

\par 
We also note that $\varphi_H^{(0)}(\xi)$ in \eqref{vph} is divergent ($\varphi_H^{(0)}(\xi) \longrightarrow \infty$) for $p_x(x) = 0$
meaning that WKB approximation is not valid around the turning points, denoted in Fig.~\ref{FIG:1} as the points with zero longitudinal 
momentum. This situation was also earlier found to be the case for graphene,\,\cite{zalip} as well as a regular Schr{\"o}dinger electron.

\subsection{General solution for the wave function}

General transport equation is 

\begin{equation}
\label{transeqn}
\frac{1}{\sqrt{2}} \, \hat{\mbb{O}}_{\,T} (x, p_y \, \vert \, \phi, \, \tau) \, \Psi_{\lambda} (x, p_y \, \vert \, \phi, \tau) -
\, \hat{\Sigma}^{(3)}_{x} (\phi)  \, \frac{\pr}{\pr x} \,  \Psi_{\lambda - 1}(x, p_y \, \vert \, \phi, \tau) = 0 \ ,
\end{equation} 

Each of the wavefunctions $\Psi_{\lambda-1} (x, p_y \, \vert \, \phi, \tau)$   and  $\Psi_{\lambda} (x, p_y \, \vert \, \phi, \tau)$ is expended over an orthonormal set of $\big \vert \mbox{\boldmath$v$}_1 \bigr>$, $ \big \vert \mbox{\boldmath$v$}_2 \bigr>$ and $\big \vert \mbox{\boldmath$v$}_3 \bigr>$ functions as:

\begin{eqnarray}
&& \Psi_{\lambda-1} (x, p_y \, \vert \, \phi, \tau) = 
\varphi_H^{(\lambda-1,1)} (x)\, \big \vert\mbox{\boldmath$v$}_1(x\,\vert\,\phi,\tau)\bigr> + 
\varphi_H^{(\lambda-1,2)} (x)\,\big \vert\mbox{\boldmath$v$}_2(x\,\vert\,\phi,\tau)\bigr>+
\varphi_H^{(\lambda-1,3)} (x)\,\big \vert\mbox{\boldmath$v$}_3(x\,\vert\,\phi,\tau)\bigr> \, , \\
\nonumber 
&& \Psi_{\lambda} (x, p_y \, \vert \, \phi, \tau) = 
\varphi_H^{(\lambda,1)} (x)\, \big \vert\mbox{\boldmath$v$}_1(x\,\vert\,\phi,\tau)\bigr> + 
\varphi_H^{(\lambda,2)} (x)\,\big \vert\mbox{\boldmath$v$}_2(x\,\vert\,\phi,\tau)\bigr> +
\varphi_H^{(\lambda,3)} (x)\,\big \vert\mbox{\boldmath$v$}_3(x\,\vert\,\phi,\tau)\bigr>
\end{eqnarray}

Once this is done, each side of Eq.~\eqref{transeqn} is multiplied by $\big \langle \mbox{\boldmath$v$}_1(x\,\vert\,\phi,\tau) \big \vert$,
$\big \langle \mbox{\boldmath$v$}_2(x\,\vert\,\phi,\tau) \big \vert$ and $\big \langle \vert\mbox{\boldmath$v$}_3(x\,\vert\,\phi,\tau) \big \vert$, leading us to the following three coupled differential equations

\begin{eqnarray}
&& \left[
\left(
\Theta + \Theta^{\star}
\right) \, \frac{\pr \, \varphi_H^{(\lambda-1,1)}}{\pr x} + \frac{\pr \, \Theta}{\pr x} \,
\left(
 \varphi_H^{(\lambda-1,1)} +  \varphi_H^{(\lambda-1,2)} 
\right) + 
\left( \Theta - \Theta^{\star} \right) \frac{ \pr \varphi_H^{(\lambda-1,2)}}{\pr x}
\right] \cos^2 \phi - \\
\nonumber
- && \left[
\left( \frac{\pr \, \Theta^{\star}}{\pr x} - \frac{\pr \, \Theta}{\pr x} \right) \varphi_H^{(\lambda-1,3)} + 
\left( \Theta^{\star} - \Theta \right) \, \frac{ \pr \varphi_H^{(\lambda-1,3)}}{\pr x}
\right] \, \frac{1}{2} \, \sin (2 \phi) + \\
\nonumber 
 + && \left[ 
\frac{\pr \, \Theta^{\star}}{\pr x} \left(
\varphi_H^{(\lambda-1,1)} +  \varphi_H^{(\lambda-1,2)} 
\right) + \Theta \, \left(
\frac{\pr \, \varphi_H^{(\lambda-1,1)}}{\pr x} - \frac{\pr \, \varphi_H^{(\lambda-1,2)}}{\pr x}
\right) + 
\Theta^{\star} \, \left(
\frac{\pr \, \varphi_H^{(\lambda-1,1)}}{\pr x} + \frac{\pr \, \varphi_H^{(\lambda-1,1)}}{\pr x}
\right)
\right] \, \cos^2 \phi = 0 \, ,
\end{eqnarray}
obtained using $\big \langle \mbox{\boldmath$v$}_1(x\,\vert\,\phi,\tau) \big \vert$,

\begin{eqnarray}
&& \left[
\left(
\Theta - \Theta^{\star}
\right) \, \frac{\pr \, \varphi_H^{(\lambda-1,1)}}{\pr x} + \frac{\pr \, \Theta}{\pr x} \,
\left(
 \varphi_H^{(\lambda-1,1)} +  \varphi_H^{(\lambda-1,2)} 
\right) + 
\left( \Theta + \Theta^{\star} \right) \frac{ \pr \varphi_H^{(\lambda-1,2)}}{\pr x}
\right] \cos^2 \phi - \\
\nonumber
- && \left[
\left( \frac{\pr \, \Theta^{\star}}{\pr x} - \frac{\pr \, \Theta}{\pr x} \right) \varphi_H^{(\lambda-1,3)} + 
\left( \Theta^{\star} - \Theta \right) \, \frac{ \pr \varphi_H^{(\lambda-1,3)}}{\pr x}
\right] \, \frac{1}{2} \, \sin (2 \phi) + \\
\nonumber 
 + && \left[ 
\frac{\pr \, \Theta^{\star}}{\pr x} \left(
\varphi_H^{(\lambda-1,1)} +  \varphi_H^{(\lambda-1,2)} 
\right) + \Theta \, \left(
\frac{\pr \, \varphi_H^{(\lambda-1,2)}}{\pr x} - \frac{\pr \, \varphi_H^{(\lambda-1,1)}}{\pr x}
\right) + 
\Theta^{\star} \, \left(
\frac{\pr \, \varphi_H^{(\lambda-1,1)}}{\pr x} + \frac{\pr \, \varphi_H^{(\lambda-1,1)}}{\pr x}
\right)
\right] \, \sin^2 \phi = \\
\nonumber
= && 4 \sqrt{2} \, \nu(x) \, \varphi_H^{(\lambda,2)} \, ,
\end{eqnarray}
when we multiplied by $\big \langle \mbox{\boldmath$v$}_2(x\,\vert\,\phi,\tau) \big \vert$, and, finally, 

\begin{eqnarray}
\frac{1}{2} \left( 
\Theta - \Theta^{\star}
\right) \, \left[
\frac{\pr \, \varphi_H^{(\lambda-1,1)}}{\pr x} - \frac{\pr \, \varphi_H^{(\lambda-1,2)}}{\pr x}
\right] \, \sin (2 \phi) = \sqrt{2} \, \nu(x) \, \varphi_H^{(\lambda,3)} \, . 
\end{eqnarray}

Here we also took into account the fact that $\big \langle \mbox{\boldmath$v$}_1(x\,\vert\,\phi,\tau) \big \vert$ is an eigenfunction of 
$\hat{\mbb{O}}_{\,T} (x, p_y \, \vert \, \phi, \, \tau)$ and

\begin{equation}
\big \langle \vert\mbox{\boldmath$v$}_{1}(x\,\vert\,\phi,\tau) \big \vert \hat{\mbb{O}}_{\,T} (x, p_y \, \vert \, \phi, \, \tau) \big \vert
\mbox{\boldmath$v$}_{2,3}(x\,\vert\,\phi,\tau)\bigr>  =  
\big \langle \mbox{\boldmath$v$}_{2.3}(x\,\vert\,\phi,\tau) \big \vert \hat{\mbb{O}}_{\,T} (x, p_y \, \vert \, \phi, \, \tau) \big \vert
\mbox{\boldmath$v$}_{1}(x\,\vert\,\phi,\tau)\bigr>  = 0 
\end{equation}
simply because transport operator $\hat{\mbb{O}}_{\,T} (x, p_y \, \vert \, \phi, \, \tau)$ is hermitian, i.e., 
$\hat{\mbb{O}}_{\,T} (x, p_y \, \vert \, \phi, \, \tau) = \hat{\mbb{O}}_{\,T}^{\dagger} (x, p_y \, \vert \, \phi, \, \tau)$.

\medskip
\par 
We are mostly interested in the results for a dice lattice 

\begin{eqnarray}
\label{m1}
&& - \left( \Theta + \Theta^{\star} \right) \, 
\left[ 
\frac{\pr \, \varphi_H^{(\lambda-1,+)}}{\pr x} + \frac{\pr \, \varphi_H^{(\lambda-1,-)}}{\pr x}
\right] -
\left[ \frac{\pr \, \Theta}{\pr x}^{\star} + \frac{\pr \, \Theta}{\pr x} 
\right] \, \varphi_H^{(\lambda-1,+)} - \\
\nonumber
&& - \left( \Theta - \Theta^{\star} \right) \, \frac{\pr \, \varphi_H^{(\lambda-1,3)}}{\pr x} + 
\left[
\frac{\pr \, \Theta^{\star}}{\pr x} - \frac{\pr \, \Theta}{\pr x}
\right] \, \varphi_H^{(\lambda-1,3)} = 0 \, ,
\end{eqnarray}

\medskip

\begin{eqnarray}
\label{m2}
&& \left( \Theta + \Theta^{\star} \right) \, 
\left[ 
\frac{\pr \, \varphi_H^{(\lambda-1,+)}}{\pr x} - \frac{\pr \, \varphi_H^{(\lambda-1,-)}}{\pr x}
\right] +
\left[ \frac{\pr \, \Theta}{\pr x}^{\star} + \frac{\pr \, \Theta}{\pr x} 
\right]  \, \varphi_H^{(\lambda-1,+)} + \\
&& + \left( \Theta - \Theta^{\star} \right) \, \frac{\pr \, \varphi_H^{(\lambda-1,3)}}{\pr x} - 
\left[
\frac{\pr \, \Theta^{\star}}{\pr x} - \frac{\pr \, \Theta}{\pr x}
\right] \, \varphi_H^{(\lambda-1,3)} = 4 \sqrt{2} \, \nu(x) \, \varphi_H^{(\lambda,2)} \, , 
\end{eqnarray}

\medskip

\begin{equation}
\label{m3}
 \left( 
\Theta - \Theta^{\star}
\right) \, 
\frac{\pr \, \varphi_H^{(\lambda-1,-)}}{\pr x}
 = \sqrt{2} \, \nu(x) \, \varphi_H^{(\lambda,3)} \, . 
\end{equation}
Equations \eqref{m1}-\eqref{m3} is employed to obtain the three components of the each consequent order $\varphi_H^{(\lambda,(1,2,3))}(x)$ 
from the previous terms  $\varphi_H^{(\lambda-1,(1,2,3))}(x)$ of expansion \eqref{expand}. Therefore, the general problem of obtaining all
the terms of the semiclassical wave function is now solved, which is the key result of our work. 

\subsection{First-order wave function and applicability of WKB approximation}

We apply first-order transport equation: 

\begin{equation}
\label{transeq12}
\hat{\mbb{O}}_{\,T} (\xi, p_y \, \vert \, \phi, \, \tau) \, \Psi_{1} (\xi, p_y \, \vert \, \phi, \tau) = 
\sqrt{2} \, \hat{\Sigma}^{(3)}_{x} (\phi)  \, \frac{\pr}{\pr \xi} \,  \Psi_{0}(\xi, p_y \, \vert \, \phi, \tau)\ ,
\end{equation} 

where $\Psi_{0}(\xi, p_y \, \vert \, \phi, \tau) = v_1 \varphi_H^{(0,1)}$.

We obtain

\begin{equation}
\left\{ 
\frac{\pr \, \Theta^{\star}(x)}{\pr x} + \frac{\pr \, \Theta(x)}{\pr x} 
\right\} \, \varphi_H^{(0)}(x) + 
2 \, \left[
 \Theta^{\star}(x) + \Theta(x)
\right] \,
\frac{\pr \,\varphi_H^{(0)}(x)}{\pr x} = 0 
\end{equation}

which is equivalent to Eq.~\eqref{mainEq} and leads us to the zero order wavefunction $\varphi_H^{(0)}(x)$.

\begin{equation}
\label{12}
\nu(x) \,  \varphi_H^{(1,2)}(x) = \frac{\sqrt{2}}{8}\, \left[ 
\frac{\pr \, \Theta^{\star}(x)}{\pr x} + \frac{\pr \, \Theta(x)}{\pr x} 
\right] \, \varphi_H^{(0)}(x)  \, , 
\end{equation}
as well as 

\begin{equation}
\label{13}
\nu(x) \,  \varphi_H^{(1,3)}(x) = \frac{\sqrt{2}}{2}\, \left[
 \Theta^{\star}(x)-  \Theta(x)
\right] 
\frac{\pr \,\varphi_H^{(0)}(x)}{\pr x} \, , 
\end{equation}

and, finally, the only remaining and most important equation to find $\varphi_H^{(1,1)}(x)$ from $\varphi_H^{(1,1)}(x)$ and 
$\varphi_H^{(1,1)}(x)$ obtained from Eqs.~\eqref{12} and \eqref{13}, correspondingly: 

\begin{eqnarray}
\label{11}
&& 2 \Theta^{+} \, \frac{\pr \,\varphi_H^{(1,1)}}{\pr x} +
\frac{\pr \,\Theta^{+}}{\pr x} \, \varphi_H^{(1,1)} +  
\frac{\pr \,\Theta^{+}}{\pr x} \, \varphi_H^{(1,2)} -
\frac{\pr \,\Theta^{-}}{\pr x} \, \varphi_H^{(1,3)} -
\Theta^{-} \, \frac{\pr \,\varphi_H^{(1,3)}}{\pr x} = 0
\, , \\
\nonumber
&& \Theta^{\pm}(x) = \Theta^{\star}(x) \pm \Theta(x) \, . 
\end{eqnarray}

Using Lagrange multiplier $\mu(x) = \tet{exp}\big[ \int f(x) dx \big]$, we can immediately solve equation \eqref{11}
for $\varphi_H^{(1,1)}(x)$ as 

 \begin{eqnarray}
\label{s11}
&& \varphi_H^{(1,1)}(x)  = \tet{exp} \left[
- \int f(x) d x
\right] \, \left\{ 
\text{const} - \int \tet{exp} \left[ 
\int f(\xi) d \xi
\right] \, g(x) d x
\right\} \, , \\
\nonumber 
&& f(x) = \frac{1}{2 \Theta^{+}(x)} \, \frac{\pr \,\Theta^{+}(x)}{\pr x}  \, , \\
\nonumber
&& g(x) = \frac{1}{2 \Theta^{+}(x)} \, \left\{
\frac{\pr \,\Theta^{+}}{\pr x} \, \varphi_H^{(1,2)} -
\frac{\pr \,\Theta^{-}}{\pr x} \, \varphi_H^{(1,3)} -
\Theta^{-} \, \frac{\pr \,\varphi_H^{(1,3)}}{\pr x}
\right\} \, .
\end{eqnarray}

Equation \eqref{s11} leads to the last remaining component of the first order wave function $\varphi_H^{(1,1)}(x)$. Thus, we have 
developed a technique to obtain the next order wave function, this concludes our effort to build WKB theory for $\alpha-\mc{T}_3$ model.

\medskip
Now, since all the wave function terms for the first order are obtained, it is reasonable to verify the applicability of WKB approximation, 
i.e., the magnitude of each consequent term of expansion \eqref{expand} is much less compared to the previous term, $\Psi_1(x) \ll 
\Psi_0(x)$ etc. 

Let us use equation \eqref{12} connecting  $\varphi_H^{(1,2)}(x)$ and  $\varphi_H^{(0)}(x)$ as an example with the most straightforward estimate procedure. We also consider  particle with a large longitudinal momentum $p_x(x) \gg p_y$ and $p_x(x) \approx \nu(x)$ which 
for a Dirac cone particle corresponds either to a high potential energy $V(x)$ or to a large initial kinetic energy $E_\gamma$ of the 
incoming electron. In this case, we can use the following approximations

\begin{eqnarray}
&& \left \vert \Theta^{\star}(x) + \Theta(x) \right \vert = \frac{2}{p_x(x)}{\nu(x)} \approx 1 \, , \\
&& \frac{\pr}{\pr x} \, \left \vert \Theta^{\star}(x) + \Theta(x) \right \vert \approx   \frac{p_y^2}{p_x(x)^2} \,
\frac{\pr p_x(x)}{\pr x} \, . 
\end{eqnarray} 

Finally, the first order term wave function term $\varphi_H^{(1,2)}(x) $ is estimated as  

\begin{eqnarray}
\label{12est}
&& \varphi_H^{(1,2)}(x) = \frac{\sqrt{2}}{8}\, \frac{1}{\nu(x)} \, \left\{ 
\frac{\pr \, \Theta^{\star}(x)}{\pr x} + \frac{\pr \, \Theta(x)}{\pr x} 
\right\} \, \varphi_H^{(0)}(x)  \approx \\
\nonumber
&& \approx \frac{1}{p_x(x)} \, \frac{p_y^2}{p_x(x)^2} \,
\frac{\pr p_x(x)}{\pr x}  \, \varphi_H^{(0)}(x)  \approx \left[ \frac{p_y}{p_x(x)^2} \right]^2 \, \frac{\pr p_x(x)}{\pr x} \,
\varphi_H^{(0)}(x)  = \left[ \frac{p_y}{p_x(x)} \right]^2 \, \left \vert \frac{\pr \lambda(x)}{\pr x} \right \vert \, 
\varphi_H^{(0)}(x)  \, , \\
\nonumber 
&&  \left \vert \frac{\pr \lambda(x)}{\pr x} \right \vert = \hbar \, \left[\frac{p_y}{p_x(x)} \right]^2 \, 
\frac{\pr p_x(x)}{\pr x} \, .
\end{eqnarray}
 Thus, we obtained a standard applicability 
criteria for a semiclassical WKB approximation in quantum mechanics of a slow-changing wavelength of the incoming particle and
condition $ p_x(x) \gg p_y$ discussed above.

\begin{figure} 
\centering
\includegraphics[width=0.5\textwidth]{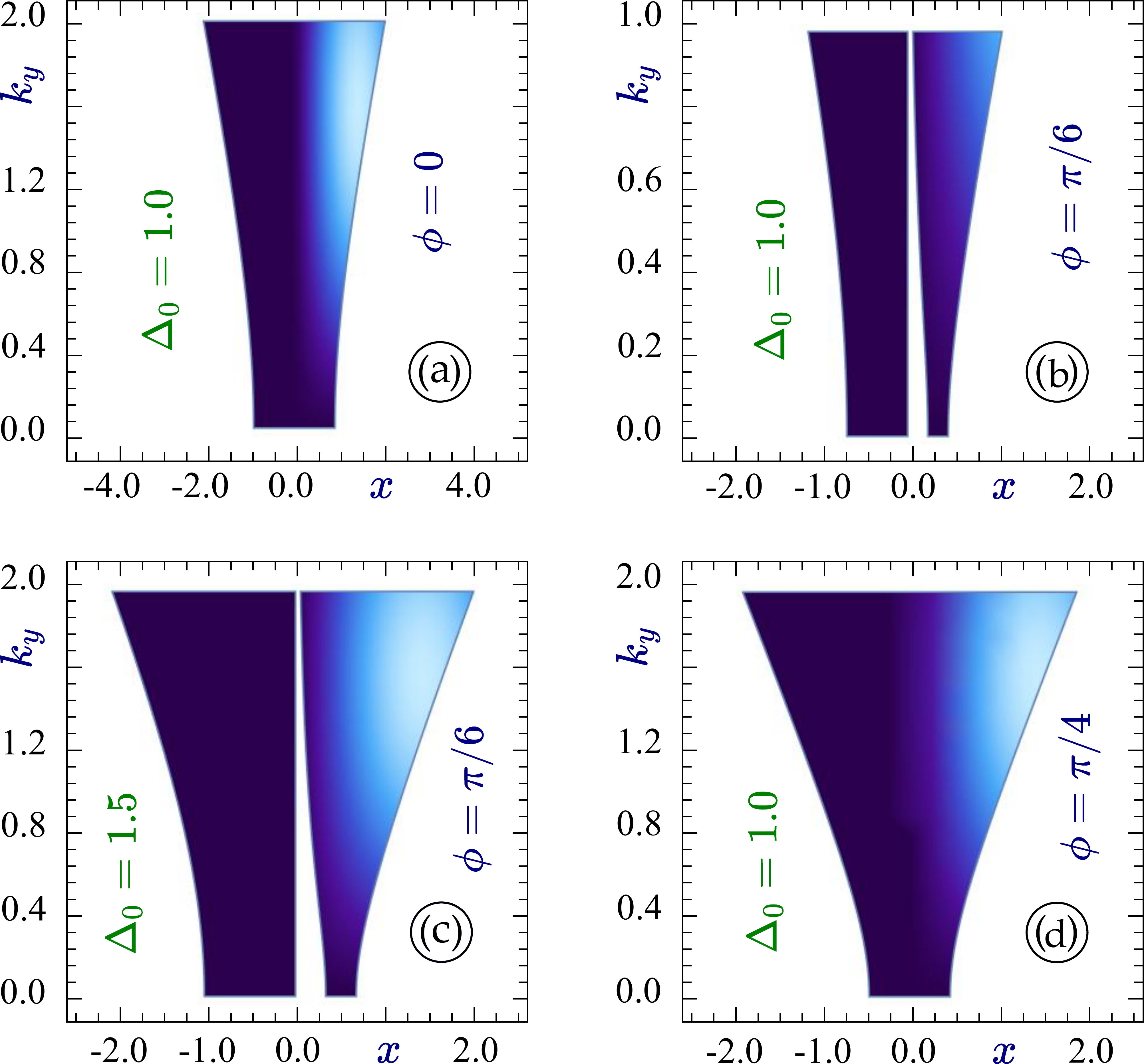}
\caption{(Color online) Classically forbidden regions (C.F.R.'s) as a function of particle's position $x$ and transverse momentum $k_y$ for an electron in gapped $\alpha-\mc{T}_3$ materials with various bandgap values moving under a linearly increasing potential $V(x) = V_0 + a x$, where $a = 1.0$ for all plots and $V_0$ was chosen to satisfy $\nu(x) = V(x)- E_\gamma = 0$ for $x=0$ in each of the considered cases. The values of the phase $\phi$ were taken $\phi = 0$ (graphene), $\phi = \pi/6$ and $\phi = \pi/4$ (a dice lattice), as labeled.}
\label{FIG:2}
\end{figure}

\begin{figure} 
\centering
\includegraphics[width=0.5\textwidth]{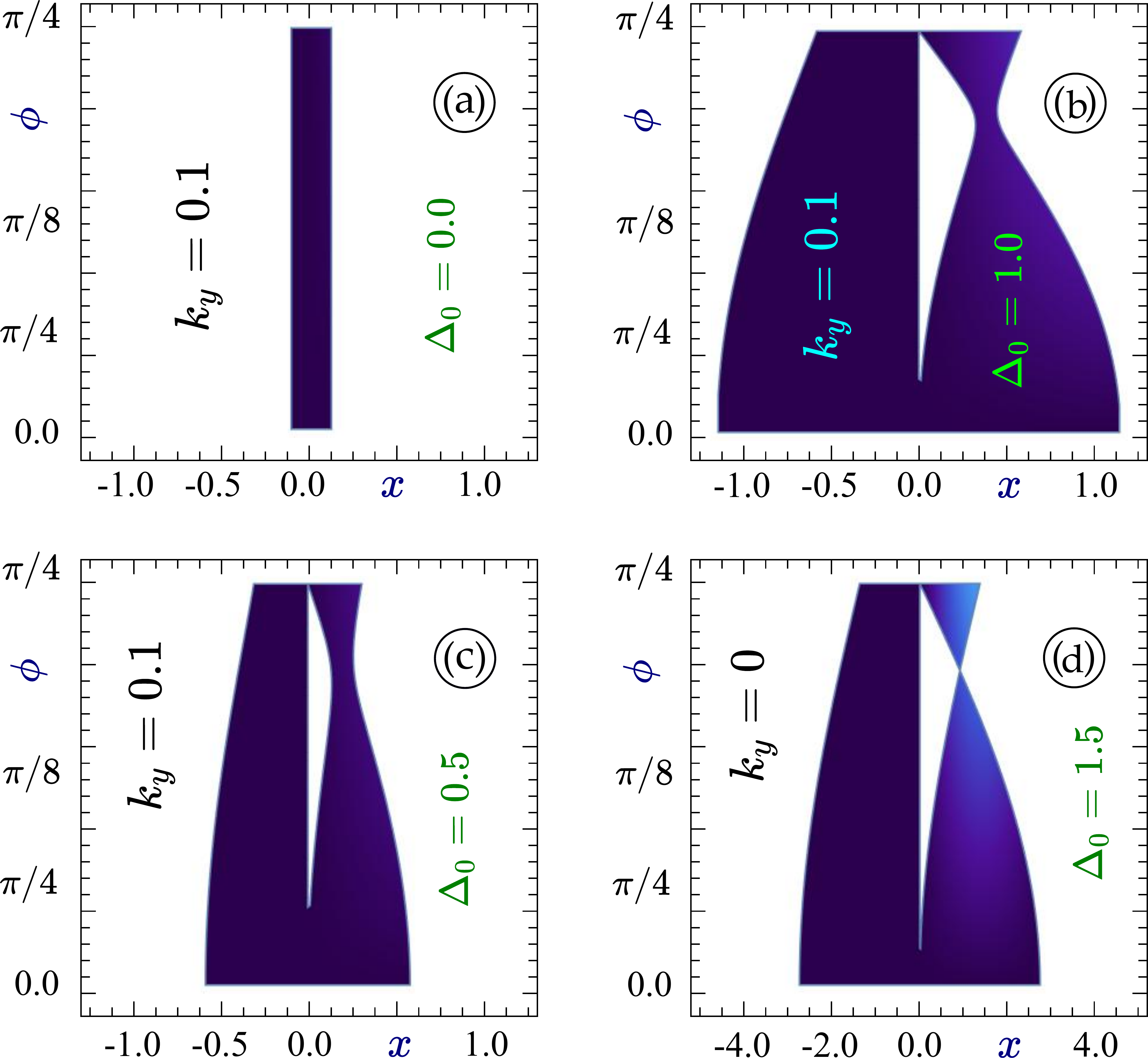}
\caption{(Color online) Classically forbidden regions (C.F.R.'s) as a function of particle's position $x$ and phase $\phi$ for an electron in gapped $\alpha-\mc{T}_3$ materials with various bandgap values moving under a linearly increasing potential $V(x) = V_0 + a x$, where $a = 1.0$ for all plots and $V_0$ was chosen to satisfy $\nu(x) = V(x)- E_\gamma = 0$ for $x=0$ in each of the considered cases. The values of the transverse momentum $k_y$ were taken $k_y = 0.1$ for panels $(a)$-$(c)$ and $k_y = 0$ for plot $(d)$.}
\label{FIG:3}
\end{figure}

\begin{figure} 
\centering
\includegraphics[width=0.5\textwidth]{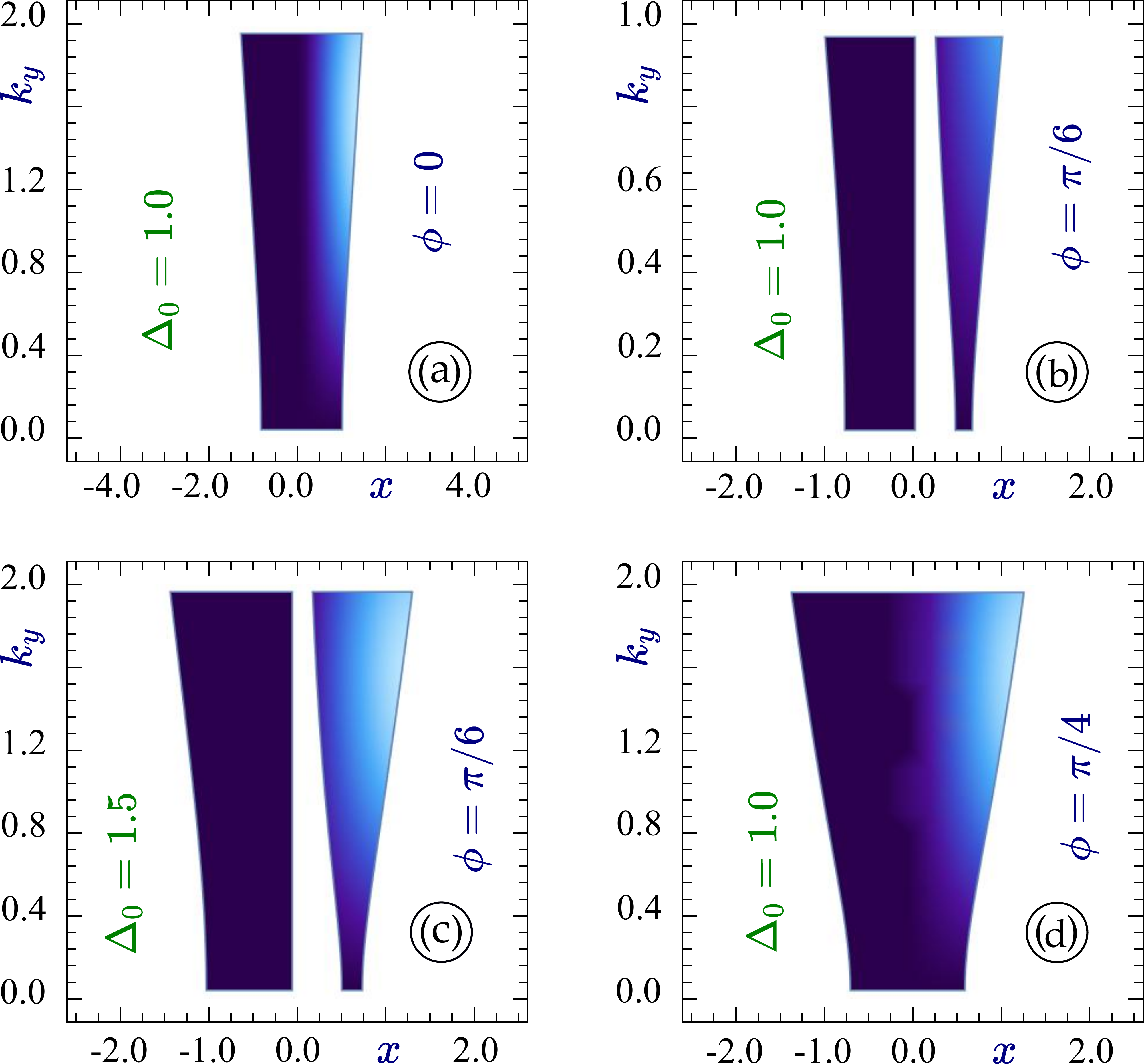}
\caption{(Color online) Classically forbidden regions (C.F.R.'s) as a function of particle's position $x$ and transverse momentum $k_y$ for an electron in gapped $\alpha-\mc{T}_3$ materials with various bandgap values moving under a linearly increasing potential $V(x) = V_0 +  a \, \text{sign}(x)\,x^2$, where $a = 1.0$ for all plots and $V_0$ was chosen to satisfy $\nu(x) = V(x)- E_\gamma = 0$ for $x=0$ in each of the considered cases. The values of the phase $\phi$ were taken $\phi = 0$ (graphene), $\phi = \pi/6$ and $\phi = \pi/4$ (a dice lattice), as labeled.}
\label{FIG:4}
\end{figure}

\begin{figure} 
\centering
\includegraphics[width=0.5\textwidth]{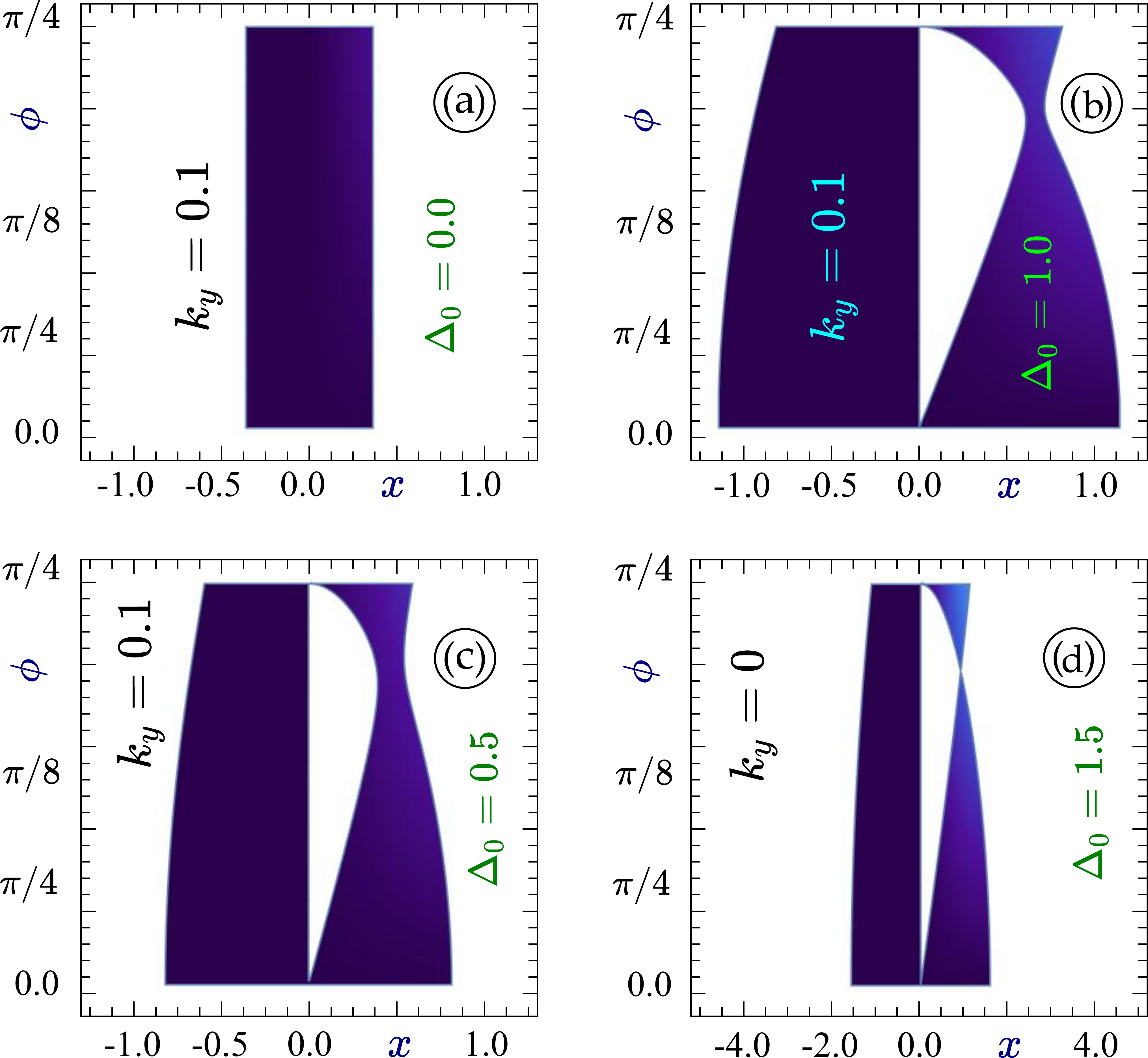}
\caption{(Color online) Classically forbidden regions (C.F.R.'s) as a function of particle's position $x$ and phase $\phi$ for an electron in gapped $\alpha-\mc{T}_3$ materials with various bandgap values moving under a linearly increasing potential $V(x) = V_0 + a \, \text{sign}(x)\,x^2$, where $a = 1.0$ for all plots and $V_0$ was chosen to satisfy $\nu(x) = V(x)- E_\gamma = 0$ for $x=0$ in each of the considered cases. The values of the transverse momentum $k_y$ were taken $k_y = 0.1$ for panels $(a)$-$(c)$ and $k_y = 0$ for plot $(d)$.}
\label{FIG:5}
\end{figure}

\section{Applications of WKB approximation}
\label{sec4} 

In this Section, we will briefly comment on how we can use the obtained approximated semiclassical wave function. The primary applications of a semiclassical theory is finding the electron transmission for non-trivial potential profiles where a direct calculation of the transmission amplitude is either impossible or too complicated, as well as studying the resonant tunneling or scattering with electron-hole transition.

If we only know action $S(x)$ which is directly related to the $x-$dependent longitudinal momentum $p_x(x) = \pr S(x)/\pr x$, it is sufficient to estimate the electron transmission through a barrier as an integral of $p_x(x)$ 

\begin{equation}
\label{Tev}
T(p_y \, \vert \, a, \Delta_0, \phi) = \tet{exp} \left[ 
- \frac{2}{\hbar} \, \int\limits_{\rm CFR} \vert p_x(x) \vert \, d x 
\right]
\end{equation}
over the so-called classically forbidden regions where $p_x(x)^2 < 0$, e.i., the longitudinal momentum is purely imaginary.\,\cite{sonin, fanwar} Such states of the particle result in a significant drop of the transmission coefficient, which is reflected in Eq.~\eqref{Tev}.

Let us consider a special case of an $\alpha-\mc{T}_3$ material with a finite energy bandgap $\Delta_0$. Such a bandgap could be achieved either by adding a dielectric substrate or irradiating our sample with an external off-resonance dressing optical field.\,\cite{Gusgap,ourpeculiar} The Hamiltonian for a gapped $\alpha-\mc{T}_3$ lattice is obtained by adding a $\phi$-dependent $\hat{\Sigma}_z(\phi)$ term 

\begin{eqnarray}
\label{h2}
&& \hat{\mc{H}}_{\Delta} (\phi) = \frac{\Delta}{2} \, \hat{\Sigma}_z(\phi) \, 
= \Delta \, \left[
\begin{array}{ccc}
\cos^2 \phi & 0 & 0 \\
0 & - \cos2 \phi & 0 \\
0 & 0 & - \sin^2 \phi
\end{array}
\right] \ , \\
\nonumber 
&& \hat{\hat{\Sigma}}_z(\phi) = -i \,
\left[\hat{\Sigma}_x(\phi), \hat{\Sigma}_y(\phi) 
\right]
\end{eqnarray}

as it was used in Ref.\,[\onlinecite{tutul1}].

\medskip
\par

The transport operator $\hat{\mbb{O}}_{\,T} (x, p_y\, \vert \, \phi, \tau)$  for Hamiltonian \eqref{h2} is modified as 

\begin{eqnarray} \label{opt-2}
&& \hat{\mbb{O}}_{\,T} (x, p_y, \Delta_0 \, \vert \, \phi, \tau) = 
\, \left[
\begin{array}{ccc}
\nu(x) + \Delta_0 \cos^2 \phi & \cos \phi \, \Xi(x)_{\mbb{S},\tau}^{-} & 0 \\
\cos \phi \, \Xi(x)_{\mbb{S},\tau}^{+} & \nu(x) - \Delta_0 \cos^2 \phi & 
\sin \phi  \, \Xi(x)_{\mbb{S},\tau}^{-} \\
0 & \sin \phi \, \Xi(x)_{\mbb{S},\tau}^{+} & \nu(x) - \Delta_0 \sin^2 \phi
\end{array}
\right] \ , \\
\nonumber
&& \Xi(x)_{\mbb{S},\tau}^{\pm} = - i \hbar \, \tau \, \frac{\pr \mbb{S}_\Delta(x)}{\pr x} \pm i p_y 
\end{eqnarray}
where $\Delta_0 \longrightarrow \Delta/V_0$. The longitudinal momentum $p_x(\xi, \Delta_0\,\vert\,\phi)$ is now given by 

\begin{equation}
\label{pixd2}
[p_x(x, \Delta_0 \, \vert \, \phi)]^2 = \nu^2(x) - p_y^2 + \frac{\Delta_0^2}{8 \,\nu(x)} \, \sin (2 \phi) \, \sin (4 \phi) 
- \frac{\Delta_0^2}{8} \left[5 + 3 \cos (4 \phi) \right] \, . 
\end{equation}

The classically forbidden regions calculated by Eq.\,\eqref{pixd2} for gapped $\alpha-\mc{T}_3$ are presented in Fig.\,\ref{FIG:2} - \ref{FIG:5}
for a linear $V(x) = V_0 + a \, x$  and non-linear $V(x) = V_0 + a \, \text{sign}(x) \, x^2$ potential profiles, correspondingly;
$a$ is determined by the magnitude of the electric field which created the potential barrier $V(x)$. We see a substantial difference
between the two types of potential which also strongly affects the electron tunneling.  We have also noted that the span of the classically forbidden regions is generally decreased for a larger phase $\phi$. 
\par
it is also important to realize that for $0 < \phi < \pi/4$ there is a pole at $\nu(x) = V(x) - E_\gamma = 0$ and the sign of the term 
$\Delta_0^2/[8 \,\nu(x)] \, \sin (2 \phi) \, \sin (4 \phi)$ would change. This means that $x=0$ must always be a boundary of a classically
inaccessible region, which we observe in all plots of Figs.~\ref{FIG:2} - \ref{FIG:5}, except for graphene or a dice lattice.

\newpage

\section{Summary and Remarks}
\label{sec5}

To summarize, this paper embodies our effort to build up a complete semiclassical description for $\alpha-\mc{T}_3$ model which describes
a wide class of Dirac cone materials with an additional flat band in their dispersions, also referred to as pseudospin-1 Dirac materials.  
Up to the present time, there has been immense and strong experimental evidence for the existence of a flat energy band in a large number of lab-fabricated materials. Any of these materials fits the theoretical $\alpha-\mc{T}_3$ model with a certain degree of precision so that our results are expected to be applicable to a plenty of existing two-dimensional lattices.

\medskip
Our development of Wentzel–Kramers–Brillouin (WKB) approximation includes finding the semiclassical action specifically for $\alpha-\mc{T}_3$ model, as well as the $x$-dependent longitudinal momentum; calculating the leading ($0^{th}$) order wave function, all of its components, their phase differences and the spatial dependence. Most importantly, we have derived a complete set of transport equations which connect any two consequent orders of the semiclassical wave function ($\Psi_\lambda(x, p_y \, \vert \, \phi, \tau)$ and $\Psi_{\lambda+1}(x, p_y \, \vert \, \phi, \tau)$). Therefore, the sought WKB wave function could be now obtained up to any required order and precision. Finally, we have solved these transport equations for the $1^{st}$ order wave function and demonstrated the conditions for the applicability of Wentzel–Kramers–Brillouin approximation for our model Hamiltonian, i.e., $\Psi_1(x, p_y \, \vert \, \phi, \tau) \ll \Psi_0(x, p_y \, \vert \, \phi, \tau)$. 

\par
Our derivations and the obtained results for $\alpha-\mc{T}_3$ appeared to be drastically different from standard WKB approximation for a Schr{\"o}dinger particle studied in Quantum Mechanics textbooks, as well as the previously obtained results for graphene.\,\cite{zalip}

\medskip
The obtained equations and the semiclassical electronic states could be further used to investigate most of the crucial electronic properties of $\alpha-\mc{T}_3$ materials. We have discussed the possible applications of the obtained wave function, primarily, studying the tunneling and transport properties of pseudospin-1 Dirac electrons in non-trivial potential profiles. Specifically, we have considered a bandgap in $\alpha-\mc{T}_3$ - a situation for which the direct computation of the wavefunction is complicated and is not even possible if the potential becomes spatially non-uniform. The calculation of the electron transmission in such cases is done by integrating the longitudinal electron momentum $\vert p_x(x) \vert$ over the so-called classically forbidden regions with imaginary momentum $p_x(x)$ ($p_x(x)^2 < 0$). We have calculated the transmission for various kind of $\alpha-\mc{T}_3$ lattices and, specifically, demonstrated that the transmission is mostly reduced for a larger phase $\phi$.

\par 
In general, our developed semiclassical theory for a pseudospin-1 Dirac electron could be employed to investigate resonant tunneling and scattering on a potential barrier, as well as to recognize trapped or localized electronic states in $\alpha-\mc{T}_3$.

\medskip
We are confident that our work  is crucial for investigating main electronic properties of a whole class of innovative low-dimensional structures.  Building a Wentzel–Kramers–Brillouin approximation has been viewed as one of the key results for each newly discovered material or a model Hamiltonian. The obtained results will undoubtedly find their applications in the follow-up research into electron tunneling and 
transport and electronic control in $\alpha-\mc{T}_3$ materials, building devices and transistors based on resonant tunneling or scattering of Dirac electrons through a custom-made potential barrier profile. Our results are also important in terms of the valleytronic applications since the low-energy states of a gapped material directly depend on the valley index, as well as many other application of these unique and innovative materials.

\begin{acknowledgements}
A.I. would like to acknowledge the funding received from TRADA-52-113, PSC-CUNY Award \# 64076-00 52. D.H. was supported by the Air Force Office of Scientific Research (AFOSR).  G.G. would like to acknowledge Grant No. FA9453-21-1-0046 from the Air Force Research Laboratory (AFRL).
\end{acknowledgements}

\bibliography{WKBbib}

\end{document}